\documentclass[sigconf, authorversion]{acmart}

\usepackage{listings}
\usepackage{multirow}

\copyrightyear{2023} 
\acmYear{2023} 
\setcopyright{acmlicensed}
\acmConference[SC-W 2023]{Workshops of The International Conference on High Performance Computing, Network, Storage, and Analysis}{November 12--17, 2023}{Denver, CO, USA}
\acmBooktitle{Workshops of The International Conference on High Performance Computing, Network, Storage, and Analysis (SC-W 2023), November 12--17, 2023, Denver, CO, USA}
\acmPrice{15.00}
\acmDOI{10.1145/3624062.3624173}
\acmISBN{979-8-4007-0785-8/23/11}

\newcommand{\markchanges}[1]{#1}

\newcommand{\camerareadyrevision}[1]{#1}
\newcommand{\additionalrevision}[1]{#1}

\begin{document}

\title[GPU Graph Processing on CXL-Based Microsecond-Latency External Memory]
{GPU Graph Processing on CXL-Based \\ Microsecond-Latency External Memory}

\author{Shintaro Sano, Yosuke Bando, Kazuhiro Hiwada, Hirotsugu Kajihara, Tomoya Suzuki, \\ Yu Nakanishi, Daisuke Taki, Akiyuki Kaneko, Tatsuo Shiozawa}
\email{{shintarou.sano, yosuke1.bando, kazuhiro.hiwada, hirotsugu.kajihara, tomoya.suzuki}@kioxia.com}
\email{{yu.nakanishi, daisuke.taki, akiyuki.kaneko, tatsuo.shiozawa}@kioxia.com}
%\email{}
\affiliation{%
  \institution{Kioxia Corporation}
  %\streetaddress{P.O. Box 1212}
  \country{Japan}
}

\renewcommand{\shortauthors}{Sano et al.}

\begin{abstract}
  In GPU graph analytics, the use of external memory such as the host DRAM and solid-state drives is a cost-effective approach to processing large graphs beyond the capacity of the GPU onboard memory.
  This paper studies the use of Compute Express Link (CXL) memory as alternative external memory for GPU graph processing in order to see if this emerging memory expansion technology enables graph processing that is as fast as using the host DRAM. % and is faster than using the SSDs.
  Through analysis and evaluation using FPGA prototypes, we show that representative GPU graph traversal algorithms involving fine-grained random access can tolerate an external memory latency of up to a few microseconds introduced by the CXL interface as well as by the underlying memory devices.  % through the PCIe Gen 4.0 x16 link.
  This insight indicates that microsecond-latency flash memory may be used as CXL memory devices to realize even more cost-effective GPU graph processing while still achieving performance close to using the host DRAM.
\end{abstract}

\begin{CCSXML}
  <ccs2012>
     <concept>
         <concept_id>10010583.10010786.10010787</concept_id>
         <concept_desc>Hardware~Analysis and design of emerging devices and systems</concept_desc>
         <concept_significance>500</concept_significance>
         </concept>
     <concept>
         <concept_id>10010583.10010786.10010809</concept_id>
         <concept_desc>Hardware~Memory and dense storage</concept_desc>
         <concept_significance>500</concept_significance>
         </concept>
   </ccs2012>
\end{CCSXML}

\ccsdesc[500]{Hardware~Analysis and design of emerging devices and systems}
\ccsdesc[500]{Hardware~Memory and dense storage}
\keywords{CXL, memory, flash, GPU, graph, latency}

% Do we need these? They were there by default

\maketitle

\section{Introduction}
\label{sec:intro}

Graphics processing units (GPUs) have become one of the most commonly-used accelerators in high-performance computing and machine learning.
In order to handle ever-growing data sizes in these applications beyond the relatively limited capacity (tens of GBs) of GPU onboard memory, the use of external memory such as the host DRAM and solid-state drives (SSDs) can be a cost-effective approach compared with pooling multiple GPUs' memory together \cite{E25,E19,E39,E37,E23,EMOGI2020,BaM41,BaM43,BaM30,BaM49,BaM34,BaM2023}.
In particular, GPU-centric external memory access methods have been shown to yield the state-of-the-art runtime performance in workloads involving on-demand, fine-grained random access such as graph analytics \cite{EMOGI2020,BaM2023}.
That is, when small pieces of data to be read next depend on the current processing results and cannot be {\it a priori} determined, it is more efficient to have the GPU initiate data requests than to have the CPU control the data flow between the GPU and external memory.

In GPU-initiated data access in graph analytics, the use of the host DRAM generally leads to faster processing speeds than SSDs (see Sections~\ref{sec:we_dont_care_file_loading_time} and \ref{sec:xlfdd_runtime} for details).
However, % the host DRAM
% (as well as other DIMM-attached memory such as Optane DCPMM) <- defer this to later discussion
% has a scaling issue meaning that increasing the capacity beyond a certain point becomes too costly.
increasing the host DRAM capacity to accommodate large graph data can be costly.
Memory expansion via Compute Express Link (CXL) \cite{CXL} is a promising alternative, as it allows load/store access to pooled memory in a cache-coherent manner over more expandable PCIe links.
% One of the major use cases of CXL is memory expansion, and it would allow the GPU to access flash memory in the same way as the host DRAM, meaning the same GPU code would work with the GPU memory, host DRAM, and CXL memory.
That being said, CXL memory introduces additional latency to the underlying memory devices (e.g., DRAM), and an added latency of one or two hundred {\it nano}seconds is shown to already have an adverse performance impact on some of CPU-based workloads \cite{Pond}.

In this paper, we are concerned about the use of CXL memory as external memory for GPU graph processing in order to see if this emerging memory expansion technology enables graph processing that is as fast as using the host DRAM.
The question we are interested in is whether GPU graph processing is tolerant to longer latency CXL introduces, and, if so, how much longer.
The latter part of the question is because, if the allowable latency is longer than the DRAM-based CXL memory latency, less expensive memory devices including low-latency flash memory may be used in place of DRAM.
In fact, our analysis indicates that, as opposed to the case of CPU workloads mentioned above, representative GPU graph traversal algorithms are latency-tolerant thanks to their massive parallelism.
The bottleneck comes from the PCIe link, which still leaves a permissible latency of a few {\it micro}seconds.

In order to back up our analysis at this early stage of CXL deployment when supporting devices are limited in availability, we use two FPGA prototypes. % complementing each other.
The first prototype is equipped with microsecond-latency flash memory, and works as a PCIe-attached storage device \cite{XLFDD}.
While it does not support the CXL interface, it nonetheless \markchanges{supports access at a smaller address alignment size} than the standard minimum unit of 512 bytes in Non-Volatile Memory Express (NVMe) SSDs, in order to serve fine-grained random read requests in graph workloads.
The second prototype is DRAM-based CXL memory, which we implement based on Intel Agilex\textregistered 7 FPGA supporting the CXL interface.
Our FPGA design features adjustable latency for the onboard DRAM, allowing us to evaluate % not only CXL-DRAM but also emulated
CXL memory with longer latency.
% We build an evaluation platform consisting of a GPU, CXL-enabled CPU (Intel Sapphire Rapids), and multiple of the CXL memory prototpyes, where the GPU user code can access the CXL memory in the same way as the host DRAM as in EMOGI.
% This XYZ(cool name for this emulator) emulates flash memory by introducing artificial latency to the FPGA onboard DRAM, which allows us to evaluate GPU graph applications on memory of varying latency.
% We derive conditions of allowable latency and
Using the first prototype, we show that \markchanges{external memory having high random read performance backed by low-latency flash memory allows us to approach host DRAM-based GPU graph processing speeds, if it supports a small address alignment size.
% albeit with a storage interface instead of CXL. 
This also confirms that the address alignment size is the primary performance factor that sets the host DRAM-based method apart from the SSD-based method, supporting the potential of CXL-based external memory that can be accessed in the same way as the host DRAM. 
}
Using the second prototype supporting the CXL interface and allowing the same GPU code to work with the host DRAM and CXL memory, we show that the runtimes on the host DRAM and CXL memory are almost identical as long as the CXL memory latency is under a certain allowable value. % derived by our analysis.
To the best of our knowledge, evaluation of GPU graph processing on CXL memory has not been reported before.
% Implemeting CXL memory equipped with microsecond-latency flash memory is left as future work.
% In addition, we use another FPGA-based prototype device \cite{XLFDD} equipped with real (non-emulated) microsecond-latency flash memory, and demonstrate close-to-DRAM graph processing speeds.
% To complement this, we use another FPGA-based device, also used in \cite{VLDB-suzuki}, that is equipped with low-latency flash memory that supports smaller access units (such as 64 and 128 bytes).

In summary, our contributions are:
\begin{itemize}
  \item We show that the performance of GPU graph processing is lenient to external memory latency, and a few microseconds may be tolerated in achieving processing speeds comparable to using the host DRAM.
  \item Using an FPGA-based external memory device equipped with microsecond-latency flash memory, we demonstrate GPU graph processing speeds close to using the host DRAM, \markchanges{confirming the importance of small address alignments, which also applies to when the GPU accesses CXL memory.}
  \item Using another FPGA device implementing CXL memory with adjustable latency, we evaluate GPU graph processing on CXL memory for the first time, and confirm that the same GPU code runs as fast as when using the host DRAM as long as the CXL memory latency is up to a few microseconds.
\end{itemize}

\section{Preliminaries}
\label{sec:prelim}

This section provides background information by explaining graph traversal on external memory, defining our performance metric, and then briefly reviewing how {\it CPU} graph traversal on low-latency flash memory was made as fast as that on the host DRAM.

\subsection{Graph Traversal on External Memory}

\begin{figure}[h]
  \centering
  \includegraphics[trim=0mm 110mm 60mm 0mm, clip, width=\linewidth]{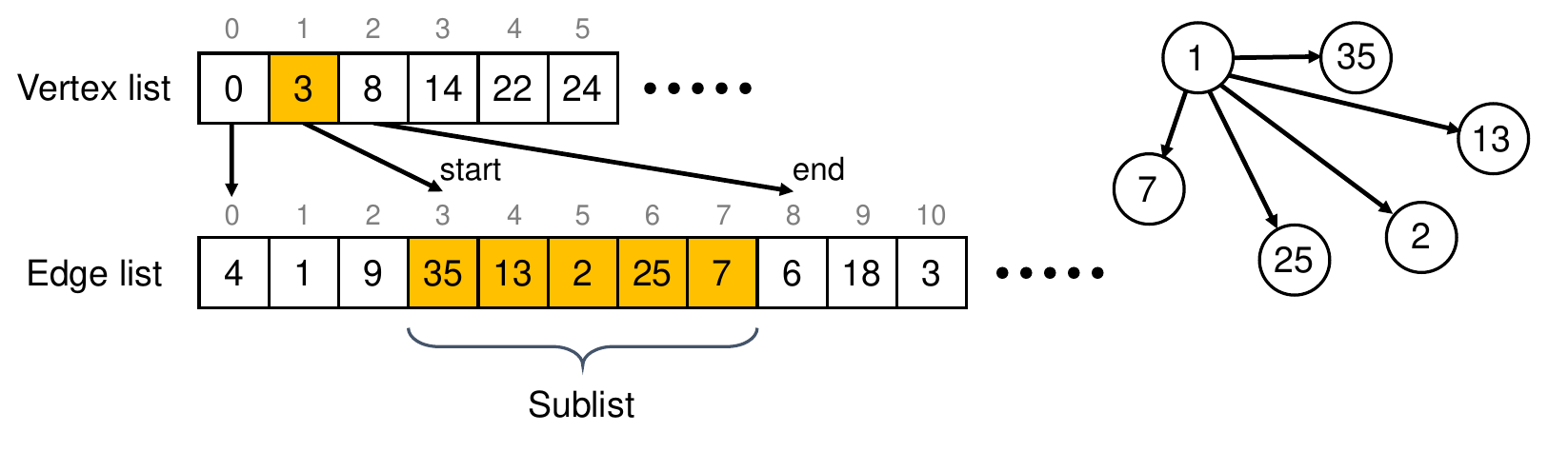}
  \caption{Compressed Sparse Row (CSR) format example.}
  \label{fig:csr}
  \Description{}
\end{figure}

A graph is represented in the commonly-used compressed sparse row (CSR) format consisting of a vertex list and an edge list as shown in Figure~\ref{fig:csr}.
% A vertex $v$ in a vertex list points to a contiguous subset in an edge list representing vertex IDs connected to $v$.
Suppose Vertex 1 points to five vertices as shown in the right-hand side of the figure.
The IDs of those five vertices appear in a contiguous subset of the edge list (called a {\it sublist} in this paper).
The start and end (exclusive) indices of this sublist is stored at Vertices 1 and 2 in the vertex list.
Since the number of edges is an order of magnitude larger than that of vertices as in Table~\ref{tbl:datasets}, the edge list is stored on the external memory. % while keeping a vertex list on the GPU memory.
When a graph is traversed, an edge sublist is read from external memory, whose size depends on the vertex's degree, which is typically a few hundred bytes on average \cite{EMOGI2020} as observed in Table~\ref{tbl:datasets}. % (8 bytes per vertex ID, and 0-degree vertices are excluded from the average).
Since vertices in this sublist determines next edge sublists to be read, access is fine-grained, random, and on-demand (cannot be determined beforehand).
% As with previous work \cite{EMOGI}, we assume the GPU code takes these data structures as input without any preprocessing.

\begin{table}[h]
  \caption{Graph datasets.} % used in evaluation. Numbers in parentheses are for the case 0-degree vertices are excluded.}
  \label{tbl:datasets}
  \begin{tabular}{cccc}
  \toprule
  % Dataset & \# of vertices & \# of edges & Ave. degrees \\
  \multirow{2}{*}{Dataset} & Num. of & Num. of edges  & Ave. degrees* \\
                           & vertices  & (edge list size) & (sublist size) \\
  \midrule
  %urand27 & 134 M (134 M) & 4.4 G & 32.0 (32.0) \\
  %kron27 & 134 M (63 M) & 4.2 G & 31.5 (67.0) \\
  %Friendster & 125 M (66 M) & 3.6 G & 28.9 (55.1) \\
  %urand27    & 134 million  & 4.4 billion & 32.0 \\
  %kron27     & 134 million  & 4.2 billion & 31.5 \\
  %Friendster & 125 million  & 3.6 billion & 28.9 \\
  %urand27    \cite{GAP2015}    & 134 mil. & 4.4 bil. (35.2 GB) & 32.0 (256.0 B) \\
  %kron27     \cite{GAP2015}    & 134 mil. & 4.2 bil. (33.6 GB) & 31.5 (252.0 B) \\
  %Friendster \cite{Friendster} & 125 mil. & 3.6 bil. (28.8 GB) & 28.9 (231.2 B) \\
  urand27    \cite{GAP2015}    & 134 mil. & 4.4 bil. (35.2 GB) & 32.0 (256.0 B) \\
  kron27     \cite{GAP2015}    & 134 mil. & 4.2 bil. (33.6 GB) & 67.0 (536.0 B) \\
  Friendster \cite{Friendster} & 125 mil. & 3.6 bil. (28.8 GB) & 55.1 (440.8 B) \\
  \bottomrule
  \multicolumn{4}{l}{\small{* 8 bytes per vertex ID. 0-degree vertices are excluded from the average.}}\\
  \end{tabular}
\end{table}

\subsection{Processing Time as Performance Metric}
\label{sec:we_dont_care_file_loading_time}

As mentioned in Section~\ref{sec:intro}, the host DRAM-based method EMOGI \cite{EMOGI2020} is generally faster than the SSD-based method BaM \cite{BaM2023} in terms of {\it graph processing time}.
If EMOGI's runtime includes the time for loading graph data onto the host DRAM from the SSDs (data loading time) in addition to the time for running the algorithm on the GPU (graph processing time), BaM is shown to be competitive for some benchmark workloads.
However, in real-world applications where one might perform more complex graph analytics, the data loading time can be negligible, making the graph processing time dominant.
Moreover, since we are interested in the use of non-volatile memory, graph data may be stored on CXL memory from the beginning without loading from SSDs.
% Therefore, we are interested in graph processing time alone in this paper.
Thus, we use graph processing time alone as a performance metric in this paper.

\subsection{Review of CPU Graph Processing Case}
\label{sec:review_cpu}

In {\it CPU} graph processing, it has previously been shown that, using microsecond-latency flash memory as external memory, processing speeds can be close to when using the host DRAM \cite{VLDB-suzuki}.
As multiple dies of microsecond-latency flash memory can support sufficient random read performance required for in-memory-class graph processing, naive external memory execution slows down not due to random read performance
\camerareadyrevision{(in input/output operations per second, or IOPS)}
of the external memory, but due to the longer latency of external memory and the CPU overhead of issuing a large number of read requests.
These issues are overcome by using lightweight context switching to hide the latency and a lightweight storage access method to reduce the CPU overhead.

\section{Analysis}
\label{sec:analysis}

This section characterizes the performance of GPU graph processing in terms of how the GPU accesses edge data.
% Ideally, 
% This section derives requirements for external memory that theoretically do not cause performance degradation from GPU graph processing where the host DRAM is used as external memory.
% The results will indicate how much the specifications may be relaxed from those of the host DRAM and will guide what memory devices may be used as external memory.

The performance characteristics of {\it GPU} graph processing is different from the {\it CPU} case reviewed in Section~\ref{sec:review_cpu}.
The difference comes from massively parallel compute resources available on the GPU and the relatively limited bandwidth of the PCIe link to the GPU.
This puts the bottleneck on the PCIe link.
In fact, both of the state-of-the-art GPU graph processing methods EMOGI (based on the host DRAM) and BaM (SSDs) achieve a data transfer rate close to the peak PCIe bandwidth.
Therefore, the performance is primarily determined by how the PCIe bandwidth to the GPU is utilized effectively.
% Either the host DRAM or multiple SSDs can provide a sufficient throughput to saturate the PCIe link between the CPU and GPU, and the GPU has sufficient computing power to process the fetched data as well as sufficient parallelism to hide the latency \cite{EMOGI2020}.
% The overhead of making read requests is not an issue either: in our evaluation environment using the storage-type FPGA prototype, by repeatedly requesting data from the same address so that the storage latency is minimal, we see 310 MIOPS using only one of streaming multiprocessors of the GPU.
% Therefore, the performance gap between the host DRAM and SSD-based methods boils down to how the PCIe bandwidth to the GPU is utilized effectively.

To examine this, we look at the runtime $t$ of a given graph traversal task for a given graph dataset in terms of how fast the GPU consumes data through the PCIe link, as
\begin{equation}
  \label{eqn:runtime}
  % t(a) = D(a)/T(a),
  % t = D/T,
  t = \frac{D}{T}
\end{equation}
where $D$ is the total data size to be read from external memory to complete the task, and $T$ is the average data throughput (in MB/sec) to the GPU.
Obviously, we would like to decrease $D$ and increase $T$ for faster execution.
% Ideally, the best performance is obtained when $D$ is equal to the edge list size $E$, and $T$ hits the PCIe bandwidth $W$.

\markchanges{
Using this equation, our analysis proceeds as follows.
Ideally, the best performance is obtained when $D$ is equal to the sum $E$ of the edge sublist sizes needed to be accessed to complete the task, and when $T$ hits the PCIe bandwidth $W$.
}
However, external memory access is done in units of a certain address alignment size coming from hardware and cache implementations, which amplifies the total data size $D$.
We will see how this amplification behaves as a function of address alignment size $a$ in Section~\ref{sec:readamp}, while the throughput $T$ will be modeled in Section~\ref{sec:throughput}. 
\markchanges{Once Equation~\ref{eqn:runtime} is characterized, Section~\ref{sec:existing} revisits existing methods EMOGI and BaM in light of Equation~\ref{eqn:runtime}.
As will be described in Section~\ref{sec:observ}, this revisit suggests opportunities for low-latency memory devices to be a cost-effective alternative to the host DRAM and be more performant than standard SSDs when used as external memory for GPU graph processing.
% Our analysis in Section~\ref{sec:observ} suggests that memory devices may have up to a few microsecond latency.
In Section~\ref{sec:latency}, we confirm that the microsecond latency allowance coming from the PCIe bottleneck is unlikely to be limited by other factors.}

\subsection{Read Amplification}
\label{sec:readamp}

\begin{figure}[h]
  \centering
  \includegraphics[trim=0mm 130mm 110mm 0mm, clip, width=\linewidth]{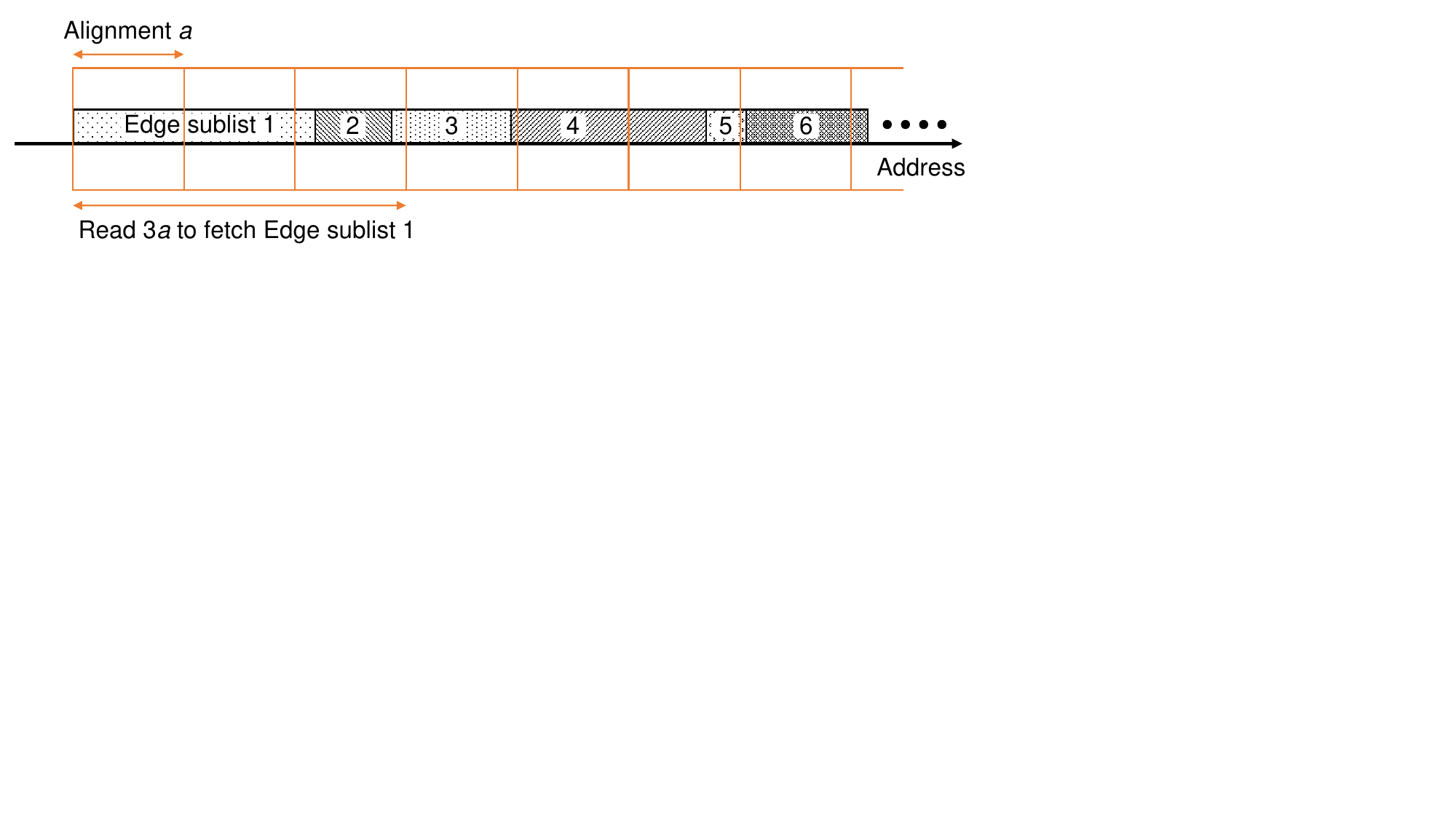}
  \caption{Aligned reads of edge sublists.}
  \label{fig:alignment}
  \Description{}
\end{figure}

% The bandwidth utilization largely depends on the ratio of the fetched data to the data actually used (i.e., read amplification factor, or RAF).
% A large RAF indicates the bandwidth is wasted by reading unused data.
When edge sublists are read from external memory at a certain address alignment size $a$, fetched data may contain some unused part of the edge list.
For example, Figure~\ref{fig:alignment} shows the situation where $3a$ bytes need to be read in order to fetch Edge sublist 1.
% If the immediately following graph traversal operation happens to require Edge sublist 2, it may be on the GPU cache, but it still contains part of Sublist 3 that may not be used soon (and may be evicted from the cache before it is referenced later).
\markchanges{Suppose the current graph traversal step requires Sublists 1 and 2 but not 3.
Then, after reading the $3a$ bytes, Sublist 2 is likely to be on the GPU cache, but the $3a$ bytes still contain part of Sublist 3 that will not be used soon (and may be evicted from the cache before it is referenced later).}
Therefore, the ratio of the fetched data to the data actually used, $D/E$, is generally greater than one.
We refer to this ratio as {\it read amplification factor}, or RAF.
In general, smaller alignments are better at reducing the RAF, given the small average edge sublist size of a few hundred bytes in graph processing.

\begin{figure}[h]
  \centering
  \includegraphics[width=0.8\linewidth]{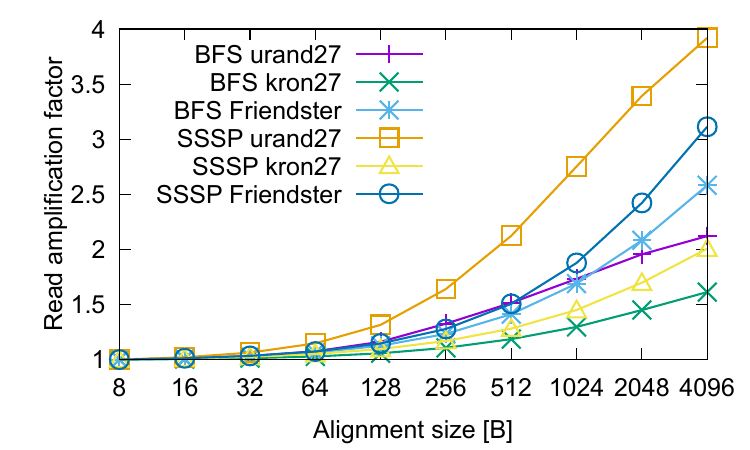}
  \caption{Read amplification for varying alignment size.}
  \label{fig:readamp}
  \Description{}
\end{figure}

To illustrate this, we ran representative graph traversal algorithms involving fine-grained random access, breadth first search (BFS) and single-source shortest path (SSSP), for varying alignment sizes and calculated the RAF.
This is CPU simulation implementing a software cache to experiment with alignment sizes without hardware constraints, but we confirmed that our RAF evaluation \markchanges{of BFS} with 512 B and 4 kB alignments match the BaM measurements well.
% Smaller RAFs mean better bandwidth utilization (less wasted data transfer).
% As we implemented a software cache, this is essentially BaM executed on the CPU and the host DRAM.
% The reason for the CPU execution is because the minimum access unit of BaM (i.e., 512 B) is determined by the NVMe SSD, and RAFs for smaller access units cannot be measured.
% We confirmed that our RAF evaluation of 512 B and 4 kB alignments match the BaM measurements well.
% We used a single cache line per warp (\textcolor{red}{need to explain warps}) as this is known to be the minimum cache size that still yields the best performance \cite{BaMPhD}.
Figure~\ref{fig:readamp} shows RAF values for the three graph datasets in Table~\ref{tbl:datasets}.
% \textcolor{red}{explain it uses a single cache line. perhaps better to explain in other words like once fetched it can be reused}
As shown, the RAFs are increasing functions of the alignment size, which can be up to 4 at 4 kB.
% Therefore, smaller access units are desirable, and as long as the GPU can saturate the PCIe bandwidth at the native cache line size of 128 B, it is expected that GPU graph processing on CXL memory may be as fast as that on host DRAM.
% Given the PCIe Gen 4.0 x16 lane link, it can be shown that CXL memory may have up to around 4 usec latency.

\subsection{Throughput}
\label{sec:throughput}

Here we examine the denominator of Equation~\ref{eqn:runtime}.
The throughput can be modeled by the following equation.
\begin{equation}
  \label{eqn:throughput}
  % T = \min\{ Sd, N_{\max} \, d/L, W \},
  T = \min \left\{ Sd, \; \frac{N_{\max}}{L} \, d, \; W \right\},
  % min{ (IOPS) x (access unit), (# outstandings)/(latency) x (access unit), (PCIe bandwidth) }
\end{equation}
where $d$ is the average data transfer size per read request, $S$ is the random read performance in IOPS of the external memory, $L$ is the average latency (including latencies of the PCIe link, CXL interface, and memory devices), $N_{\max}$ is the maximum number of outstanding (in other words, in-flight or concurrent) requests that can be issued through the PCIe link, and $W$ is the PCIe bandwidth.
The first term in the $\min$ operation trivially states that the throughput is the product of the IOPS and the data size per IO, but it is capped by the PCIe bandwidth $W$ in the third term.
The second term introduces an additional limit imposed by Little's Law stating that (by adapting it to our case) the data size passing through the link at any given time instance ($N$ concurrent data transfers of size $d$) is equal to the product of the throughput and latency: %$Na = TL$.
\begin{equation}
  \label{eqn:littles_law}
  Nd = TL.
\end{equation}
This means that the throughput is capped as $T = Nd/L \le N_{\max}\,d/L$.
% Note that this limit is imposed for memory (host DRAM or CXL) access but not for storage access, because in the latter case, the correspondence between a request and a response is taken care of by the storage interface rather than by PCIe.
\markchanges{Note that this limit by PCIe is imposed for memory (host DRAM or CXL) access but not for storage access.
In the storage case, the limit comes from the queue depth of the storage interface, which is typically much larger than $N_{\max}$ when multiple drives are used.}

\begin{figure}[h]
  \centering
  \includegraphics[width=0.9\linewidth]{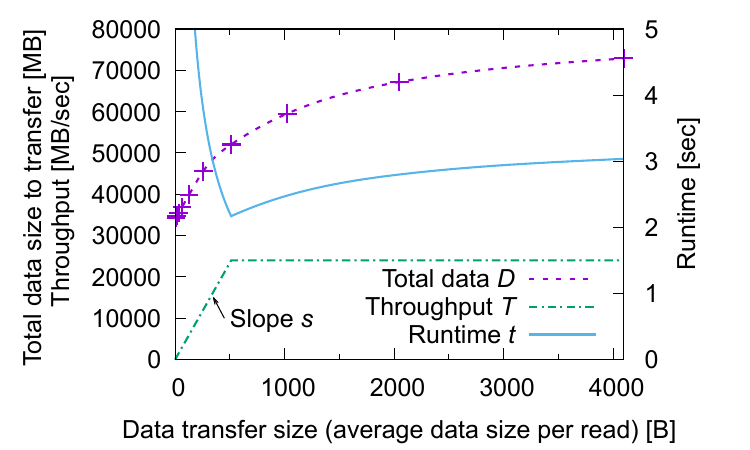}
  \caption{Runtime as a function of data transfer size.}
  \label{fig:analysis_runtime}
  \Description{}
\end{figure}

To put Equation~\ref{eqn:throughput} into context, consider a PCIe Gen 4.0 x16 link supported by modern GPUs.
\markchanges{Then, $N_{\max} = 768$ due to the PCIe specification,}
and $W = 24{,}000$ MB/sec, for which we use an effective bandwidth rather than the theoretical value of 31,500 MB/sec.
Now we suppose our external memory has $S = 100$ MIOPS (collectively, if comprised of multiple devices) and $L = 16$ usec.
\markchanges{These numbers are just for the sake of example.} % to illustrate Equation~\ref{eqn:throughput}.}
Then, Equations~\ref{eqn:throughput} becomes
\begin{equation}
  \label{eqn:throughput_with_values}
  T = \min\{ 100 \, d, \, 48 \, d, \, 24{,}000 \},
\end{equation}
which is plotted as the bottom dotted line in Figure~\ref{fig:analysis_runtime}
\markchanges{(the other two lines will be explained in Section~\ref{sec:bam})}.
We assume that the IOPS $S$ and average latency $L$ do not depend on the transfer size $d$, which is reasonable for flash-based external memory as long as the transfer size is under a certain size for which the device is optimized.
For instance, typical SSDs are optimized for 4 kB access, and reading smaller bytes does not significantly increase the random read performance due to its internal page size and error correction size.
\markchanges{Similar tendency can be observed for drives optimized for smaller sizes.}
Under this assumption, the plot linearly increases until it hits the bandwidth limit, and its slope $s$ is given as
\begin{equation}
  \label{eqn:slope}
  s = \min \left\{ S, \; \frac{N_{\max}}{L} \right\},
\end{equation}
which is 48 in the above example.

\subsection{Study of Existing Methods}
\label{sec:existing}

In order to have a shorter runtime $t = D/T$, one wishes to decrease the total data size $D$ (or equivalently, RAF) and increase the throughput $T$.
The former can be done by using a smaller alignment size $a$ while the latter by using a larger data transfer size $d$.
%Data transfer can be made at a multiple of alignment size $a$.
We look at the state-of-the-art methods, EMOGI and BaM, in light of these objectives and see how they achieve optimal runtimes within their respective constraints.
A PCIe Gen 4.0 x16 link is assumed.

\subsubsection{EMOGI}
\label{sec:emogi}

EMOGI uses the host DRAM as external memory.
It employs zero-copy access to the host DRAM, meaning that data is fetched from the host DRAM without copying it to the GPU memory.
The data access is performed in the same way as to the GPU memory, and requests are issued at a multiple of 32 B up to the GPU's hardware cache line size of 128 B \cite{EMOGI2020}.
Therefore, we have alignment size $a = 32$ B due to the GPU architecture, and the average data transfer size depends on the workload and how EMOGI cleverly issues 32 B reads so that the GPU merges them into a larger size when an edge sublist spans multiple of 32 B alignments \cite{CUDACoalescing}.
From their evaluation \cite{EMOGI2020}, we assume the distribution of 32, 64, 94, 128-B accesses to be 20\%, 20\%, 20\%, 40\%, which translates to the average transfer size of $d_{\rm EMOGI} = 0.2 \times 32 + 0.2 \times 64 + 0.2 \times 96 + 0.4 \times 128 = 89.6$ B.
This is a conservative estimate \markchanges{(i.e., the worst case among the distributions reported in \cite{EMOGI2020})}, as 128-B reads are more dominant in many workloads.

As for the RAF, $a = 32$ B alignment % is small enough relative to the average edge sublist size of a few hundred bytes, and
is close to optimal as can be seen in Figure~\ref{fig:readamp}: smaller alignments will have a diminishing return.

Regarding the throughput, $d_{\rm EMOGI} = 89.6$ B is sufficient in maximizing the throughput to saturate the PCIe bandwidth.
Since the IOPS of the host DRAM-based external memory is excessively high, the slope $s$ of the throughput in Equation~\ref{eqn:slope} is limited by the latency, which is around 1.2 usec as seen from the GPU (measured in Section~\ref{sec:cxl_character} as shown in Figure~\ref{fig:cxl_xl_latency}).
This latency is still short enough because $s \, d_{\rm EMOGI} = (768/1.2) \times 89.6 = 57{,}344$ MB/s, which is greater than the PCIe bandwidth of $W =$ 24,000 MB/sec.

% In summary, given the PCIe bottleneck, EMOGI should achieve the shortest possible runtime among external memory-based GPU graph traversal methods.

\subsubsection{BaM}
\label{sec:bam}

BaM uses SSDs as external memory.
As BaM implements a software cache on the GPU memory and reads data at a cache line granularity, we have $d = a$.
In this case, we can plot both of $D$ and $T$, and hence $t$ as well, as a function of transfer size $d$.
Figure~\ref{fig:analysis_runtime} plots examples of them.
% The total data size $D$ is taken from BFS for urand27 dataset and smoothly interpolated.
The plot of the total data size $D$ smoothly interpolates the data points taken from BFS for urand27 dataset.
Note that it shows the raw data size in bytes along the linear horizontal axis in Figure~\ref{fig:analysis_runtime} as opposed to Figure~\ref{fig:readamp} showing RAF along the $\log_2$ axis.
By dividing this $D$ by the example throughput profile $T$ described in Section~\ref{sec:throughput}, the solid line in Figure~\ref{fig:analysis_runtime} shows the theoretically-expected runtime $t$ of BFS algorithm for urand27 dataset for varying transfer sizes $d$.
It is clear from this plot that the best (shortest) runtime is obtained at the minimum transfer size that still fully utilizes the bandwidth $W$.
That is, the optimal transfer size $d_{\rm opt}$ satisfies $s \, d_{\rm opt} = W$. %, leading to:
% \begin{equation}
%   \label{eqn:best_size}
%   d_{\rm opt} = \frac{W}{\min \left\{ S, \; \frac{N_{\max}}{L} \right\}}.
% \end{equation}
%
% BaM uses four of Intel P5800X SSDs totaling $S = 6$ MIOPS with $L = O(10)$ usec latency.
% Putting these into Equation~\ref{eqn:best_size} by assuming $L = 10$, we have
% \begin{equation}
%   \label{eqn:best_size_bam}
%   d_{\rm BaM} = \frac{24{,}000}{\min\{6, 76.8\}} \approx 4 \; {\rm kB},
% \end{equation}
% which is indeed the cache line size mainly used in their evaluation.
Since BaM uses four of Intel P5800X SSDs totaling $S = 6$ MIOPS, and $s = S$ as this is storage access, the optimal size is given as $d_{\rm BaM} = W/S = 24{,}000/6 \approx 4$ kB, which is indeed the cache line size mainly used in their evaluation.

Here, as seen from the denominator of the equation $W/S$, the IOPS is the limiting factor, requiring BaM to use a large data transfer size.
% The latency will not be a bottleneck even if we use $L = 100$ usec.
This suggests that one might be able to achieve faster runtimes by using higher-IOPS memory devices and smaller transfer/alignment sizes.
% If latency is considered fixed, the runtime will, under the assumption in this analysis, improve until the IOPS reaches 76.8 IOPS, in which case the optimal alignment is 24000/76.8 = 312 B.

\subsection{Observations}
\label{sec:observ}

From the analyses of the two methods above, we can see that EMOGI's faster runtime than BaM can be primarily explained by the fact that EMOGI's alignment of 32 B is smaller than that of BaM (typically 4 kB).
\markchanges{Both methods maximize the throughput as $T = W$ in Equation~\ref{eqn:runtime}, and therefore the difference comes from the total data size $D$, which prefers smaller alignments.}
Of course, there is a good reason why BaM chooses a large alignment size for the storage in use as explained in Section~\ref{sec:bam}, but it also suggests a possibility of faster processing by using a smaller alignment size supported by high-IOPS memory devices.

In the meantime, Section~\ref{sec:emogi} shows that EMOGI's average transfer size is more than sufficient to fully utilize the PCIe bandwidth, indicating that we might be able to relax the specifications of external memory from those of the host DRAM.
More specifically, in order to satisfy $s \, d_{\rm EMOGI} \ge W$, we have
\begin{equation}
  \min \left\{ S, \; \frac{768}{L} \right\} \times 89.6 \ge 24{,}000.
\end{equation}
This becomes $S \ge 268$ MIOPS and $L \le 2.87$ usec.
Therefore, an additional latency of a few microseconds, introduced by the CXL interface and the underlying memory devices, may be tolerated.
\markchanges{While the IOPS requirement is rather high, it is feasible by bundling multiple high-IOPS (tens of MIOPS) devices together.}
% The overhead of making read requests is not an issue either: in our evaluation environment using the storage-type FPGA prototype, by repeatedly requesting data from the same address so that the storage latency is minimal, we see 310 MIOPS using only one of streaming multiprocessors of the GPU.

\markchanges{
In summary, our observations are as follows.
On the condition that we have sufficient random read performance,}
%\begin{itemize}
%  \item {\bf Observation 1:} A smaller address alignment size is better.
%  \item {\bf Observation 2:} The allowable latency is a few microseconds.
%\end{itemize}
\vspace{1mm}

\markchanges{{\bf Observation 1:} A smaller address alignment size is better.}

\markchanges{{\bf Observation 2:} The allowable latency is a few microseconds.}

\subsection{Latency}
\label{sec:latency}

So far, our latency $L$ is limited by the allowable number $N_{\max}$ of outstanding requests in the PCIe specification (256 for PCIe Gen 3.0 and 768 for Gen 4.0 and 5.0).
% Here we examine whether the latency may be capped by other factors that limit the number of concurrency.
% Our conclusion is that currently the number of PCIe tags is what limits the allowable latency.
\markchanges{However, there are also other factors having their own concurrency limits.
Here we examine them and confirm that, currently, the strictest limit comes from PCIe.} 
% In PCIe Gen 6.0 with $N_{\max} = 15{,}360$, this situation is likely to change, which means even longer latency may be tolerated in the near future.

\subsubsection{Traversal Algorithm}

Typically, BFS-like graph traversal is massively parallelizable.
Table~\ref{tbl:frontier} shows how many vertices are being visited (i.e., frontier) at each depth of the search in BFS for urand27 dataset.
Most depths have more than tens of thousands of vertices that can be processed independently, indicating that the algorithm itself does not limit concurrency.
Some depths have smaller frontiers, but they contribute little to the overall runtime.
\begin{table}[ht]
  \caption{Example numbers of vertices per traversal depth.}
  \label{tbl:frontier}
  \begin{tabular}{rr}
  \toprule
  Depth & Number of vertices \\
  \midrule
  1 &          31 \\
  2 &         984 \\
  3 &      31,252 \\
  4 &     995,253 \\
  5 &  28,130,066 \\
  6 & 104,931,066 \\
  7 &     129,075 \\
  \bottomrule
  \end{tabular}
\end{table}

\subsubsection{GPU}
There is an upper limit to the GPU concurrency, \markchanges{but it is much larger than $N_{\max} = 768$, and thus the GPU will not be a limiting factor.}
% The GPU we use has 64 {\it streaming multiprocessors} (SMs) each with 48 {\it warps} (refer to \cite{CUDAProgrammingModel,CUDACoalescing} for the details of SM and warp).
% As each warp consists of 32 GPU threads that all execute the same instructions and collectively issue read requests, the warp is considered a unit of concurrency.
\markchanges{A group of GPU threads that execute the same instructions is called a {\it warp} \cite{CUDAwarps}, which is considered a unit of concurrency.
The GPU we use has 3,072 warps.
Fewer warps may actually run depending on the workload due to the limitations of other resources such as GPU registers.
Yet, in our BFS execution, we find that 2,048 warps are running, which is still larger than $N_{\max}$.}
% If all of them run concurrently, the allowable latency can be computed using Equation~\ref{eqn:littles_law} as $L = Nd_{\rm EMOGI}/T = 2{,}048 \times 89.6 / 24{,}000 = 7.64$ usec.
% To see if this is the case, while running EMOGI, we used the GPU clock to record the times before and after the edge list access on the host DRAM. % as in Listing~\ref{lst:time}.
% \lstset{language=C}
% \begin{lstlisting}[xleftmargin=15pt,frame=single,basicstyle=\small,caption={Measuring Edge Access Time \label{lst:time}}]
%   time_before = clock();
%   vertex_id = edge_list[index];
%   time_after = clock();
% \end{lstlisting}
% The time difference was 8.46 usec on average (with a standard deviation of 2.17 usec), which roughly matches the allowable latency computed above.
% $N = TL/d_{\rm EMOGI} = 24{,}000 \times 8.46 / 89.6 = 2{,}266$, which roughly matches the number of warps 2048.
% Therefore, the GPU can handle a sufficient number of outstanding reads, permitting a latency of around 8 usec, which is less strict than what the PCIe Gen 4.0 and 5.0 link impose.

\subsubsection{CXL Interface}
\label{sec:analysis_cxl}

The CXL specification itself is unlikely to limit the number of outstanding reads, as 16 tag bits are available (65,536 outstanding requests) \cite{CXLSpecIntel}.
There is a possibility that CXL memory devices do not fully utilize them depending on their implementations, leaving a stricter latency allowance.
Moreover, as the CXL data transfer size is 64 B, larger read requests from the GPU have to be split, consuming more tags.
That said, our assumption is that CXL memory devices coming to the market in the near future will support a sufficient number of outstanding requests.
% Therefore, as will be detailed in Section~\ref{sec:eval_cxl}, we deal with the concurrency limitation of our prototype by downgrading the GPU link to PCIe Gen 3.0 with $N_{\max} = 256$, so that the PCIe link limits the concurrency.

\section{Evaluation}

As CXL-enabled flash memory devices are not available yet, we use two FPGA prototypes to support the analysis presented in Section~\ref{sec:analysis}.
The first prototype, XLFDD, is a storage device equipped with low-latency flash memory reported elsewhere \cite{XLFDD}.
% \markchanges{We use this high-IOPS drives to show that a smaller alignment is indeed effective as pointed out by {\bf Observation 1} in Setion~\ref{sec:observ},}
% and demonstrate that it is possible to use microsecond-latency flash memory as external memory to approach the speed of the host DRAM-based method. % EMOGI over PCIe Gen 4.0 x16 link.
%
The second prototype is a DRAM-based CXL memory device with adjustable latency.
% While we have to downgrade the GPU PCIe link to Gen 3.0 x16 due to the prototype limitations,
% It allows us to evaluate the GPU graph processing on CXL memory for the first time, and to confirm the permissible latency imposed by the number $N_{\max}$ of outstanding requests of the PCIe link.
% \markchanges{In support of {\bf Observation 2},}
% we show that GPU graph traversals on CXL memory with a latency of up to a few microsecond can be as fast as that on the host DRAM.
\markchanges{The two prototypes respectively demonstrate {\bf Observation 1} and {\bf Observation 2} made in Section~\ref{sec:observ}.}

We use graph datasets listed in Table~\ref{tbl:datasets} including two synthetic graphs, uniform random graph (urand27) and Kronecker graph (kron27) having $2^{27}$ vertices \cite{GAP2015}, and a real-world graph Friendster \cite{Friendster}.
% The parameters of kron follows Graph500\cite{Graph500} specification.
% The degree distributions of the graphs are shown in Figure~\ref{fig:degree-distribution}.
% Numbers in the graph names denote $\log_2$ of the number of vertices. % in the graph.
We run BFS and SSSP as representative graph traversals involving fine-grained random access.

\subsection{Evaluation on Low-Latency Flash Memory}
\label{sec:eval_xlfdd}

We first show evaluation using XLFDD,
\markchanges{a high-IOPS device supporting a small alignment of 16 B backed by low-latency flash memory.
In support of {\bf Observation 1}, we show that this small alignment enables much faster GPU graph processing than BaM with a 4 kB alignment, and}
demonstrate % the possibility of using low-latency flash memory as external memory to achieve
runtime performance close to using the host DRAM.

\subsubsection{Implementation}

\markchanges{Our implementation is conceptually similar to BaM in the sense that the GPU controls the storage devices directly without CPU intervention, but there are some differences coming from the use of XLFDDs instead of NVMe SSDs.}

To explain XLFDD briefly, it is a PCIe-attached SSD equipped with low-latency flash chips with a latency of under 5 usec and with an FPGA implementing the storage controller.
% Unlike standard NVMe SSDs,
It implements a lightweight storage interface so that it can serve fine-grained accesses at up to 11 MIOPS. % , which makes it suitable external memory for graph analytics \cite{VLDB-suzuki}.
% Importantly for GPU graph processing,
It supports a 16 B alignment, % which is much smaller than the typical minimum size of 512 B of standard SSDs,
while the transfer size can be any multiple of 16 B up to 2 kB.
This transfer size flexibility allows us to read a large edge sublist in one request without splitting it into \markchanges{the GPU cache line size of 128 B as would happen in the memory (host DRAM or CXL memory) access case}.
This makes the average transfer size $d$ close to the average edge sublist size (256 B in urand27 and more in the other datasets), further relaxing the requirements for the external memory as
%\begin{equation}
%  \min \left\{ S, \; \frac{768}{L} \right\} \times 256 \ge 24{,}000,
%\end{equation}
$S \times 256 \ge 24{,}000$, leading to $S \ge 93.75$ MIOPS. % and $L \le 8.192$ usec.

% In NVMe, the CPU writes a read request to a ring buffer on the host memory (DRAM) called submission queue (SQ), and writes to the SQ doorbell register ({\it ring the SQ doorbell}) in the SSD to notify of the request.
% The SSD reads the requested data and transfers it to the data buffer on the host memory.
% The SSD writes the completion of the request to the completion queue (CQ).
% The CPU detects the completion and rings the CQ doorbell to notify the SSD that the CQ entry has been consumed.
% The typical data transfer alignment of NVMe SSDs is 512 B both for the address and transfer size.
% In other words, 
% the data transfer size and its start address are specified as a multiple of 512 B.
%\begin{figure}[t]
%  \centering
%  \includegraphics[width=\columnwidth]{figs/NVMe-XLFDD-crop.pdf}
%  \vspace*{-3mm}
%  \caption{SSD interfaces of (a) NVMe and (b) XLFDD. % in case that completion is done by checking the value in the data buffer.
%  }
%  \label{fig:NVMe-XLFDD}
%\end{figure}

Table~\ref{tbl:system_xlfdd} summarizes our evaluation environment equipped with XLFDDs.
With 16 drives, the system well supports the required random read speed of 93.75 MIOPS.
\markchanges{To evaluate BaM, we replace XLFDDs with NVMe SSDs that collectively offer 6-MIOPS random read performance to match the number used in \cite{BaM2023}.}

\begin{table}[h]
  \caption{System for evaluation on low-latency flash.}
  \label{tbl:system_xlfdd}
  %\centering
  \begin{tabular}{cl}
  \toprule
   & \multicolumn{1}{c}{Specifications} \\
  \midrule
  CPU & Intel Xeon Gold 6336Y (single socket)\\
  %\hline
  DRAM & DDR4 3200 MHz 128 GB (16 GB $\times$ 8 ch.) \\
  %\hline
  GPU & NVIDIA RTX A5000, GDDR6 24GB, PCIe 4.0 x16\\
  %\hline
  \multirow{2}{*}{SSD} & 16 of XLFDD (PCIe 3.0 x4)\\
                       & \markchanges{4 of KIOXIA FL6 800 GB NVMe (PCIe 4.0 x4)}\\
  %\hline
  OS & Ubuntu 20.04.3 LTS, Linux kernel 5.4.0\\
  %\hline
  S/W & NVIDIA Driver 495.29.05, CUDA 11.5\\
  \bottomrule
  \end{tabular}
\end{table}

As with BaM, we place submission queues (SQs) and data buffers in the base address register (BAR) section of the GPU memory % as illustrated in Figure~\ref{fig:GPU-BAR}
in order to control storage devices directly from the GPU.
Note that we do not have completion queues \cite{VLDB-suzuki}.
By memory-mapping BARs, the SQs and data buffers are accessible from XLFDDs.
% In case of using CQ for the completion, we place CQ in the BAR as well.
% (think it's okay not to mention CQ here. If CQ needs to be mentioned later, just say CQ can be implemented if needed)
% Note that we need to use GPUs that have a sufficient size for BAR including NVIDIA V100, A100, and NVIDIA RTX 30-series with resizable BARs.
% In our case, 256 MiB BAR of RTX A5000 GPU is used.
%\begin{figure}[t]
%  \centering
%  \includegraphics[width=0.7\columnwidth]{figs/GPU_BAR-crop.pdf}
%  \caption{Direct transaction between the GPU and XLFDDs.}
%  \label{fig:GPU-BAR}
%\end{figure}

Our graph processing software for XLFDD is similar to BaM. % but is not exactly the same due to the differences in the storage interface.
% The major difference is that, because we use a much smaller alignment size than BaM's, we do not implement software caches on the GPU memory, and instead directly access XLFDDs for simplicity.
\markchanges{
The major difference is that we do not implement software caches on the GPU memory, and instead directly access XLFDDs for simplicity.
Because we use a much smaller alignment size than BaM's, caches do not reduce the RAF much, and therefore this simplification has a minimal impact on performance.}

\subsubsection{Runtime Comparison}
\label{sec:xlfdd_runtime}

\begin{figure}[h]
  \centering
  \includegraphics[trim=0 3 0 10,clip,width=0.8\columnwidth]{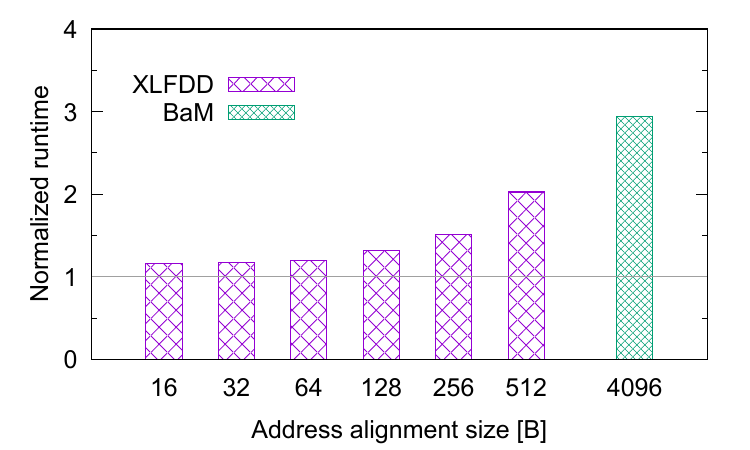}
  \caption{Runtimes of BFS for urand27 dataset on XLFDD with varying alignment sizes, along with the runtime of BaM, normalized by that of EMOGI.}
  \label{fig:xlfdd_runtime_alignment}
\end{figure}

With this system, we show that \markchanges{a small alignment size leads to higher performance ({\bf Observation 1}), and GPU graph processing speeds on low-latency flash memory can approach those on the host DRAM.
We run our software on XLFDDs, BaM on the NVMe SSDs, and EMOGI on the host DRAM. % , to execute GPU graph algorithms BFS and SSSP.
Figure~\ref{fig:xlfdd_runtime_alignment} shows the runtimes of BFS for urand27 dataset on XLFDD where we vary the address alignment size.
The runtimes are normalized by that of EMOGI, and the normalized runtime of BaM with a 4 kB alignment is also shown for comparison.
The plots demonstrate faster execution with smaller alignments, and at an alignment of 16 or 32 B, it approaches the speed on the host DRAM.}

\begin{figure}[h]
  \centering
  \includegraphics[trim=70 8 70 8,clip,width=0.49\columnwidth]{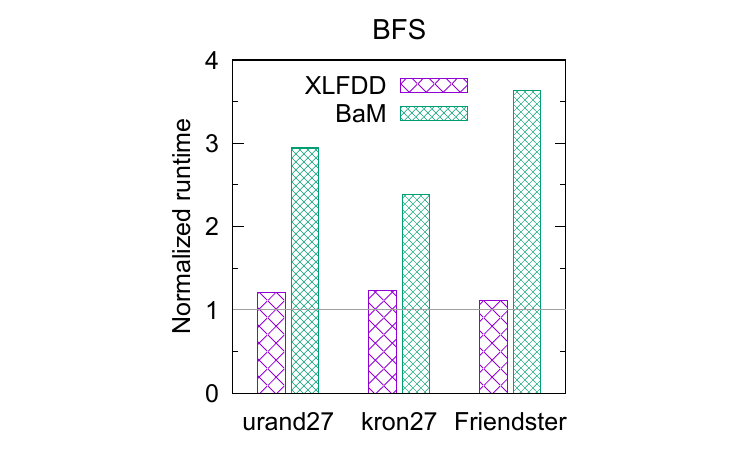}
  \includegraphics[trim=70 8 70 8,clip,width=0.49\columnwidth]{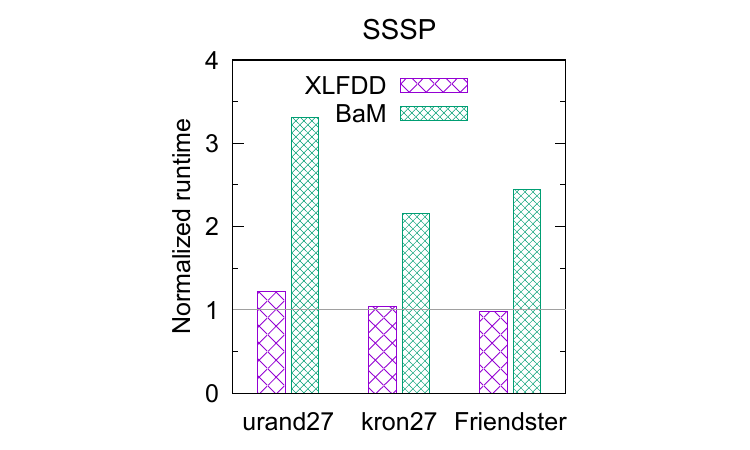}
  \caption{Runtimes of BFS (left) and SSSP (right) on XLFDD and BaM, normalized by those of EMOGI on the host DRAM.
  %\textcolor{red}{Figure is a placeholder for now. Will be updated.}
  }
  \label{fig:xlfdd_runtime_comparison}
\end{figure}

\markchanges{Figure~\ref{fig:xlfdd_runtime_comparison} compares the normalized runtimes of XLFDD and BaM for all the pairs of the algorithms and datasets, where XLFDD uses a 16 B alignment.
% As shown, GPU graph processing on XLFDD is faster than BaM thanks to the smaller alignment backed by high-IOPS drives equipped with low-latency flash memory.
The runtimes of XLFDD are much closer to those of EMOGI (1.13 times longer on average, where the geometric mean is taken over all the six pairs) than those of BaM (2.76 times longer).}

\vspace*{-3.2mm}
% XLFDD is faster than BaM by
% BFS urand 2.44, kron 1.94, friend 2.31, geomean 2.22
% for SSSP, we can't directly compare XLFDD and BaM
% as XLFDD uses 4 byte vertex while BaM uses 8 byte
%
% XLFDD is slower than EMOGI by
% BFS  urand 1.21, kron 1.23, friend 1.12, geomean 1.19
% SSSP urand 1.22, kron 1.04, friend 0.98, geomean 1.07
% geomean of six results: 1.13
%
% BaM is slower than EMOGI by
% BFS  urand 2.94, kron 2.38, friend 3.63, geomean 2.94
% SSSP urand 3.31, kron 2.16, friend 2.44, geomean 2.59
% geomean of six results: 2.76

\subsection{Evaluation on CXL Memory}
\label{sec:eval_cxl}

% Next, we evaluate GPU graph processing on CXL memory to show that the host DRAM-based method runs equally fast when CXL memory is used as external memory in place of the host DRAM, as long as the latency of the CXL memory is under a few microseconds \markchanges{as observed in {\bf Observation 2}.}

Next, we show evaluation using a DRAM-based CXL memory prototype with adjustable latency.
\camerareadyrevision{It implements the CXL.mem protocol,}
allowing us to evaluate the GPU graph processing on CXL memory for the first time, and to confirm the permissible latency imposed by the number $N_{\max}$ of outstanding requests of the PCIe link.
In support of {\bf Observation 2}, we show that GPU graph traversals on CXL memory with a latency of up to a few microsecond can be as fast as that on the host DRAM.

\subsubsection{Implementation}
\label{sec:cxl_imple}

\begin{figure}[h]
  \centering
  \includegraphics[trim=40mm 30mm 40mm 5mm, clip, width=\columnwidth]{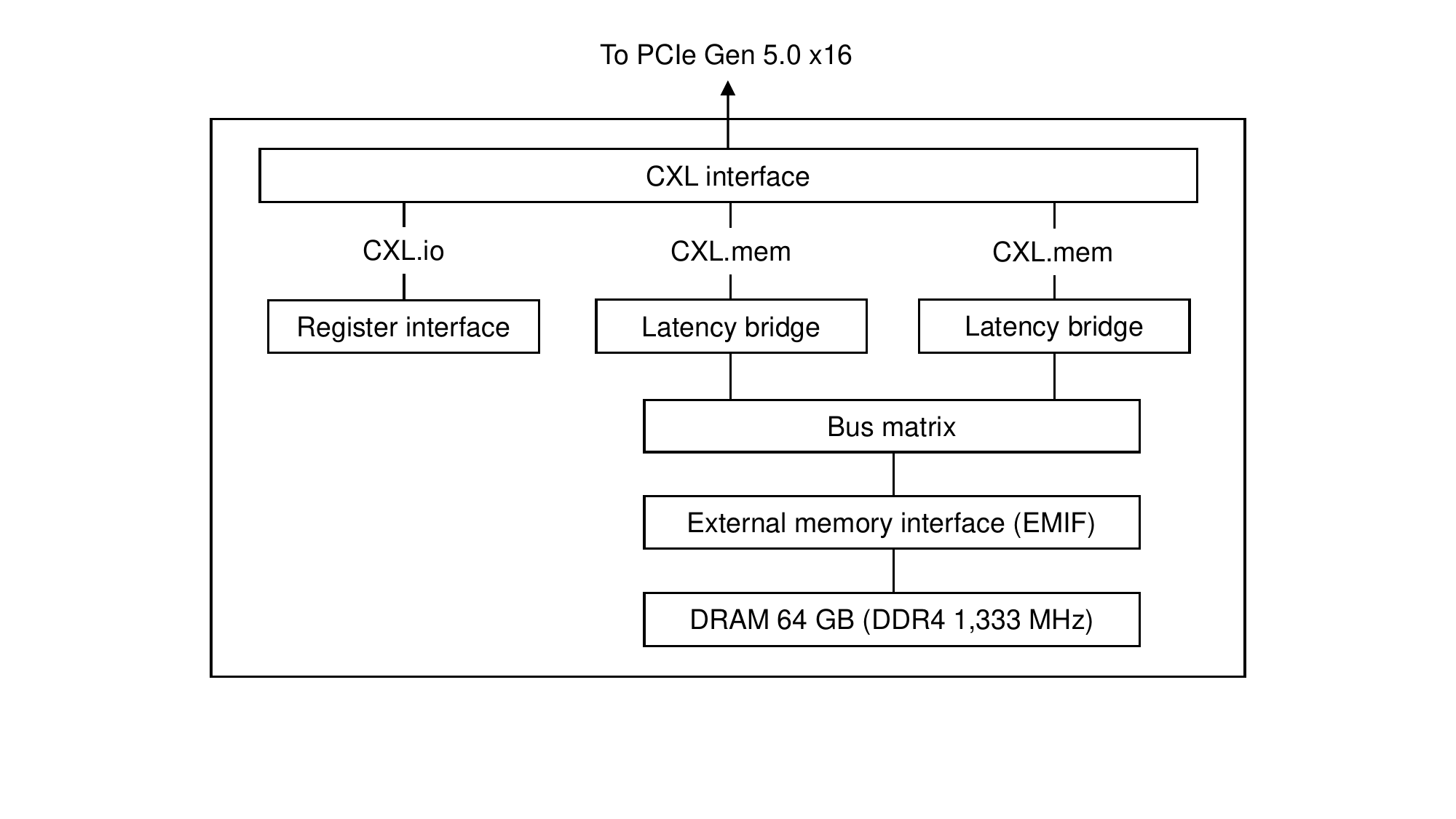}
  \caption{CXL memory prototype with adjustable latency.}
  \label{fig:CXL-XL}
\end{figure}

\markchanges{We execute EMOGI on latency-adjustable CXL memory instead of on the host DRAM.}
We implement CXL memory based on Intel Agilex\textregistered 7 FPGA as also used in other existing works \cite{DemystifyCXL,Elastic}.
Figure~\ref{fig:CXL-XL} shows the block diagram. % of the CXL memory.
% The latency bridges introduce additional latency to the FPGA onboard DRAM to emulate flash, and the behaviors of the latency bridges can be controlled by setting registers via CXL.io.
% We configure the CXL interface as Type 3 device for memory expansion.
The CXL interface has two \camerareadyrevision{instances of} CXL.mem each connecting to latency bridges that we designed to introduce additional latency to the onboard DRAM
\camerareadyrevision{(see Appendix~\ref{sec:latency_bridge} for details)}. % to emulate memory devices with longer latency.
The behaviors of the latency bridges can be controlled by setting registers via CXL.io.
Due to the limitation of the current FPGA board, the onboard DRAM can be accessed only through a single channel %, and the two slices access it
by interleaving through the bus matrix, which limits the throughput that this CXL memory prototype can support per device.

\begin{figure}[t]
  \centering
  \includegraphics[trim=30mm 100mm 30mm 0mm, clip, width=\columnwidth]{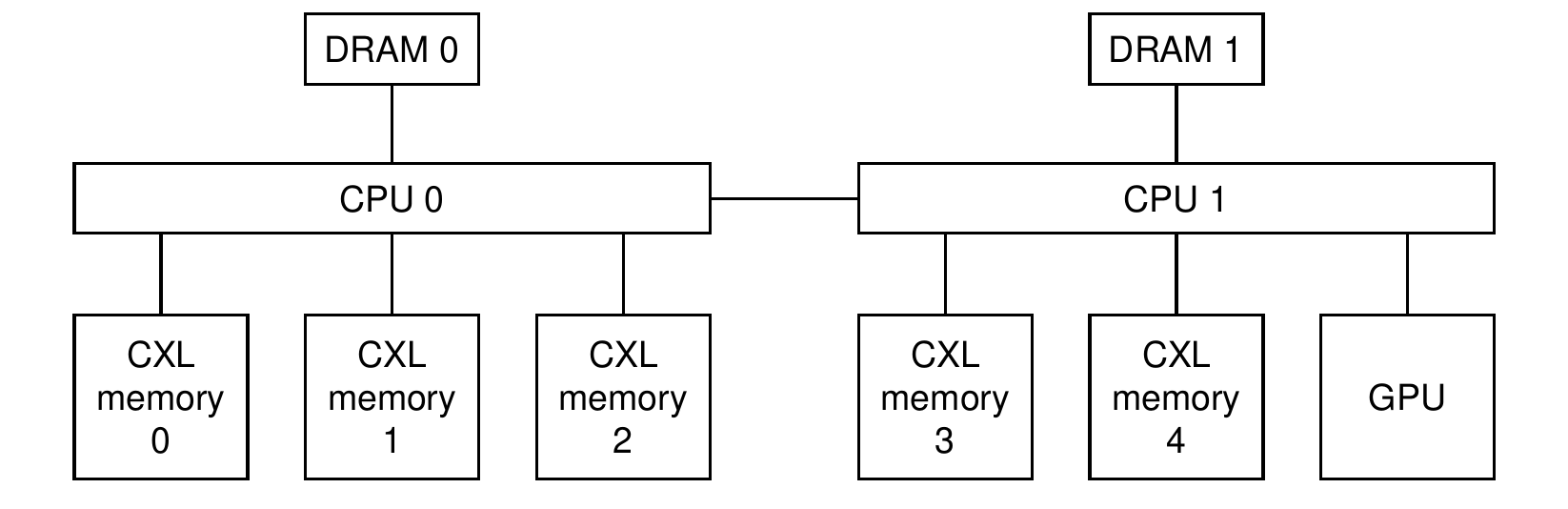}
  \caption{Connectivity of CPUs, GPU, and CXL memory.}
  \label{fig:system_cxl_xl}
\end{figure}

Table~\ref{tbl:system_cxl_xl} summarizes our evaluation system equipped with multiple of this CXL memory devices.
Figure~\ref{fig:system_cxl_xl} illustrates the connectivity between the CPUs, GPU, and CXL memory devices.
In our dual-socket system, the GPU is attached to CPU 1.

\begin{table}[t]
  \caption{System for evaluation on CXL memory.}
  \label{tbl:system_cxl_xl}
  \begin{tabular}{cl}
  \toprule
   & \multicolumn{1}{c}{Specifications} \\
  \midrule
  CPU 0/1 & Intel Sapphire Rapids* (dual socket) \\
  % CPU 0/1 & \camerareadyrevision{Intel Xeon Silver 4410Y} (dual socket) \\
  DRAM 0 & DDR5 4800 MHz 192 GB (32 GB $\times$ 6 ch.) \\
  DRAM 1 & DDR5 4800 MHz 32 GB (32 GB $\times$ 1 ch.) \\
  GPU    & NVIDIA RTX A5000, GDDR6 24 GB, PCIe 4.0 x16 \\
  CXL    & 5 of Intel Agilex\textregistered 7 FPGA I-Series Dev. Kit \\
  OS     & Fedora 34, Linux kernel 5.4.0 \\
  S/W    & NVIDIA Driver 530.30.02, CUDA 12.1 \\
  \bottomrule
  \multicolumn{2}{l}{\small{* Evaluation based on Intel reference designs and pre-production}} \\[-1mm]
  \multicolumn{2}{l}{\small{4th Gen Intel Xeon Scalable processors.}}
  \end{tabular}
\end{table}

\camerareadyrevision{It is easy to set up CXL memory to be accessible from the GPU.}
% \camerareadyrevision{In order to set up CXL memory to be accessible from the GPU, }
% In order to access CXL memory from the GPU,
One can use \texttt{set\_mempolicy()} to specify the NUMA node ID corresponding to the CXL memory device.
For instance, as explained in \cite{EMOGI2020}, \texttt{cudaMallocManaged()} can be used to allocate memory on the host DRAM for zero-copy access.
The CXL equivalent can be done by calling \texttt{set\_mempolicy()} before \texttt{cudaMallocManaged()}.
Once done, the graph processing code of EMOGI works on the CXL memory without any modification.
\additionalrevision{The GPU performs zero-copy access in the same way as does to the host DRAM, and the CPU translates it into CXL access.}
% \lstset{language=C}
% \begin{lstlisting}
% set_mempolicy(MPOL_BIND, node3);
% cudaMallocManaged(&mem_device, size);
% memcpy(mem_device, mem_host);
% set_mempolicy(MPOL_DEFAULT);
% \end{lstlisting}

\subsubsection{System Performance Characterization}
\label{sec:cxl_character}

Before running GPU graph processing, we conduct some microbenchmarks to characterize the system performance \markchanges{in terms of latency, throughput, random read performance, and the number of outstanding requests.}

\begin{figure}[h]
  \centering
  \includegraphics[trim=0 30 0 30,clip,width=0.95\columnwidth]{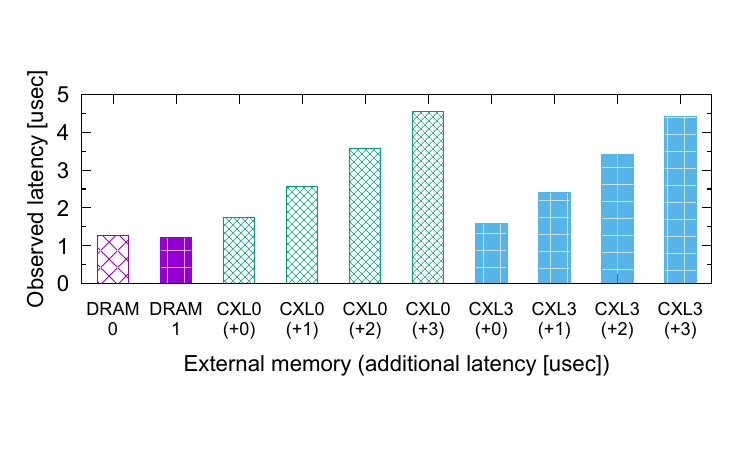}
  \caption{Measured latency of host DRAM and CXL memory as seen from the GPU.}
  \label{fig:cxl_xl_latency}
\end{figure}

First, we measure the latency introduced by the CXL memory by running pointer chasing on the GPU to access external memory \camerareadyrevision{(see Appendix~\ref{sec:pointer_chasing} for details)}.
The results are shown in Figure~\ref{fig:cxl_xl_latency}.
For the CXL memory, we vary the additional latency %, which is shown in the parentheses under the CXL memory IDs.
as shown in the parentheses.
The observed latency becomes longer as our FPGA latency bridge adds more latency as expected.
The GPU sees a latency of around 1+ usec going through the PCIe link to the host DRAM as also reported in \cite{EMOGI2020}, and the CXL DRAM introduces an additional latency of 0.5 usec. % according to our observation.
Access to external memory devices connected to the same CPU as the GPU (DRAM 1 and CXL 3 as in Figure~\ref{fig:system_cxl_xl}, which are solid-filled in Figure~\ref{fig:cxl_xl_latency}) sees marginally shorter latencies than their counterparts connected to the other CPU (DRAM 0 and CXL 0).
\camerareadyrevision{Note that the latencies reported here are those observed from the GPU, and therefore they are longer than those reported in the literature studying CXL memory access from the CPU \cite{Pond,TPP}.}

\begin{figure}[h]
  \centering
  \includegraphics[trim=0 0 10 0,clip,width=0.8\columnwidth]{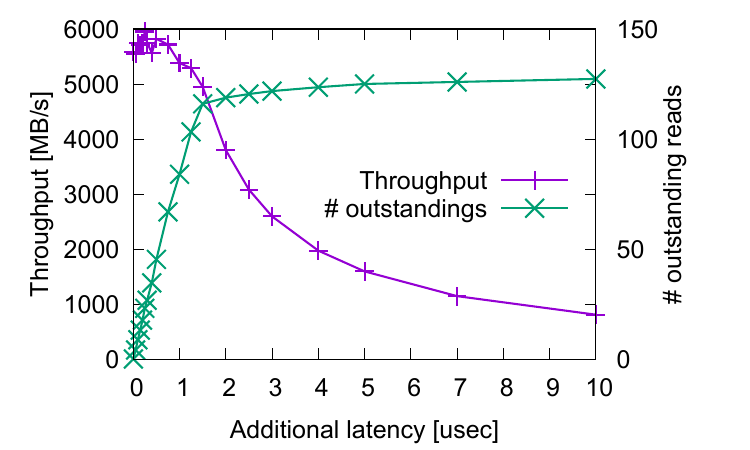}
  \caption{Bandwidth and the number of outstanding reads of the CXL memory prototype for varying additional latency.}
  \label{fig:cxl_xl_bandwidth}
\end{figure}

Next, we have the CPU (not the GPU) issue random read requests to the CXL memory prototype using the CXL access size of $d_{\rm CXL} = 64$ B, and plot the observed throughput $T_{\rm CXL}$ for varying latency as shown in Figure~\ref{fig:cxl_xl_bandwidth}.
We use the subscript \textrm{CXL} to clarify that we are referring to the CXL memory characteristics rather than the PCIe link between the CPU and GPU.
From the plot we can see that the throughput is capped at around 5,700 MB/sec due to the single-channel DRAM as mentioned above. %, which means that the maximium random read performance is 89 MIOPS ($= 5{,}700 / d_{\rm CXL}$).
The decrease in the throughput for longer latency indicates it is limited by the maximum number of outstanding requests of the CXL prototype, although the CXL specification itself permits 65536 of them as mentioned in Section~\ref{sec:analysis_cxl}.
From Equation~\ref{eqn:littles_law}, the number $N_{\rm CXL}$ of concurrent requests for a given latency $L_{\rm CXL}$ can be computed as $N_{\rm CXL} = T_{\rm CXL} L_{\rm CXL} / d_{\rm CXL}$, which is also plotted in the same figure.
This implies that the maximum number of outstanding requests that the current Intel Agilex\textregistered 7 FPGA can handle is 128.
Because a 128 B or 96 B read from the GPU through PCIe is split into two 64 B reads at the CXL level, 
\camerareadyrevision{the number of requests for the CXL memory can double}.
% one CXL memory device can handle at least 64 (= 128/2) outstanding requests from the GPUs. % (assuming much fewer 32 B and 64 B reads from the GPU).
\camerareadyrevision{Thus, our CXL memory prototype can handle 64 (= 128/2) outstanding requests from the GPUs.}
While this number is a current limitation that we expect to be lifted in the near future, in order to use this prototype to test the scenario where the concurrency bottleneck is in the PCIe link to the GPU, we downgrade the PCIe link to Gen 3.0 and use five of the CXL memory devices (which is the maximum number of devices we are able to operate in the server), such that the maximum number of outstanding requests that the CXL memory devices can collectively handle, % which is 640 ($= 128 \times 5$), is more than twice as large as that of PCIe ($N_{\max} = 256$). % for Gen 3.0).
\camerareadyrevision{which is 320 ($= 64 \times 5$), is larger than that of PCIe} \additionalrevision{Gen 3.0} ($N_{\max} = 256$).

With PCIe Gen 3.0 x16 link for the GPU, the effective bandwidth is halved as $W = 12{,}000$ MB/sec, and the requirements for external memory becomes \markchanges{$S = W/d_{\rm EMOGI} = 12{,}000/89.6 = 134$} MIOPS and $L = N_{\max} \, d_{\rm EMOGI}/W = 256 \times 89.6 / 12{,}000 = 1.91$ usec.
The halved bandwidth can be saturated by five CXL memory devices when the additional latency is less than 3 usec at which the per-device throughput is around 2,500 MB/sec as shown in Figure~\ref{fig:cxl_xl_bandwidth}.
% The IOPS requirement can also be satisfied by 5 devices since, at an additional latency of 3 usec, each device has a 64 B random read speed of 2,500/64 = 39 MIOPS, and in order to serve up to 128 B reads, the five devices can support $39 \times 5 / 2 = 97$ MIOPS.

\begin{figure*}[t]
  \centering
  \includegraphics[trim=0 1 0 6,clip,width=\columnwidth]{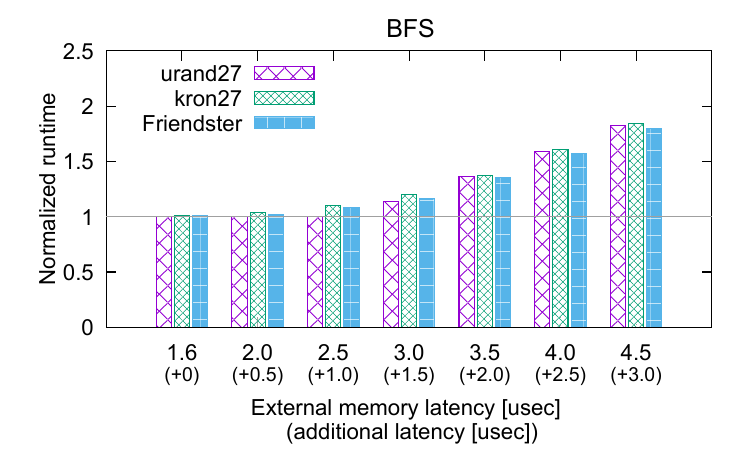}
  \includegraphics[trim=0 1 0 6,clip,width=\columnwidth]{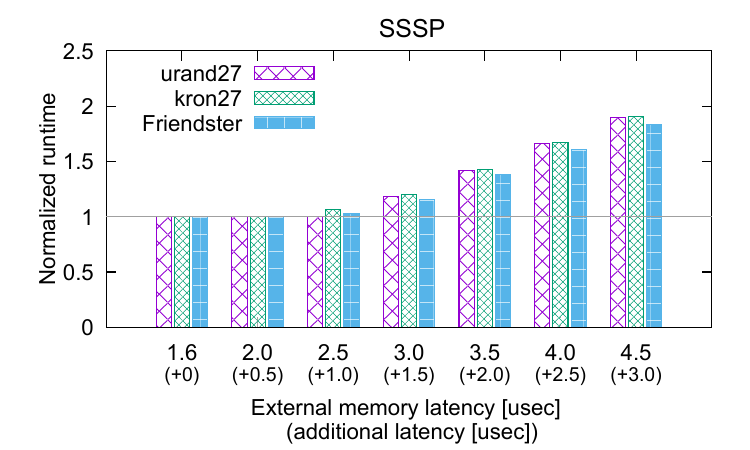}
  \caption{Runtimes of BFS (left) and SSSP (right) on CXL memory with varying latency, normalized by those on host DRAM.}
  \label{fig:cxl_xl_runtime}
\end{figure*}

\subsubsection{Runtime Comparison}
\markchanges{Here we show that GPU graph processing on CXL memory is as fast as that on the host DRAM, as long as the CXL memory latency is up to a few microseconds ({\bf Observation 2}).}
We execute BFS and SSSP on the GPU using either the host DRAM or CXL memory as external memory over PCIe Gen 3.0 x16 link to the GPU.
\markchanges{For each algorithm, the same EMOGI code is used for both the host DRAM and CXL memory.}
We vary the additional latency of the CXL memory from 0 to 3 usec. %, as 3 usec is the maximum additional latency that can still saturate the PCIe Gen 3.0 x16 bandwidth as described above.
For each graph data set, the runtimes on the CXL memory are normalized by that using the host DRAM, and plotted in Figure~\ref{fig:cxl_xl_runtime}.
The latencies written above the parenthesized additional latencies are those seen from the GPU derived according to Figure~\ref{fig:cxl_xl_latency}.
As long as the CXL memory latency from the GPU is under around 2 usec ($\approx 1.91$ usec as calculated in Section~\ref{sec:cxl_character}), we see that the runtime on CXL memory is almost identical to that on the host DRAM as expected.
% Similar observations can be made for the SSSP algorithm as in Figure~\ref{fig:cxl_xl_sssp_runtime}.

%\begin{figure}[t]
%  \centering
%  \includegraphics[width=\columnwidth]{figs/runtime_cxl_bfs.pdf}
%  \caption{BFS runtimes on the CXL memory with varying latency, normalized by runtimes on the host DRAM.}
%  \label{fig:cxl_xl_bfs_runtime}
%\end{figure}
%
%\begin{figure}[t]
%  \centering
%  \includegraphics[width=\columnwidth]{figs/runtime_cxl_sssp.pdf}
%  \caption{SSSP runtimes on the CXL memory with varying latency, normalized by runtimes on the host DRAM.}
%  \label{fig:cxl_xl_sssp_runtime}
%\end{figure}

\section{Discussion}

This section discusses limitations to our work and points not considered in the paper, all of which present future research avenues.

\vspace{0.5mm}
\noindent
{\bf Prototype limitations:}
Our evaluation using the two prototypes has both demonstrated runtime performance close to using the host DRAM in GPU graph processing on microsecond-latency external memory.
However, it has fallen short of full demonstration of our target:
% the one involving all of the CXL interface, real microsecond-latency memory devices, and PCIe Gen 4.0 x16 link to the GPU.
the one involving both the CXL interface and real microsecond-latency memory devices.
The limitations of our work come from our preliminary CXL memory implementation.
% It does not implement real low-latency flash memory, and while it does emulate longer latency, it does not model all the aspects of flash memory behaviors.
It is emulated memory with longer latency, which does not necessarily model all the aspects of real devices.
Nonetheless, we believe our flash-based (although non-CXL) prototype complements it, so that our evaluation using both prototypes jointly provides insight into how a given system %, current or future one,
may achieve host DRAM-like performance. % in GPU graph processing on CXL-based microsecond-latency external memory.
Even though the PCIe generations each double the bandwidth, the GPU and memory performance also increases accordingly.
Thus, it is likely that the PCIe link to the GPU will continue to be the bottleneck,
% and the same principle as derived in Section~\ref{sec:analysis} will apply in the foreseeable future.
and our analysis will apply in the foreseeable future.

Our analysis indicates that the full demonstration mentioned above would permit a latency of around 3 usec as described in Section~\ref{sec:observ}.
While this is still shorter than the low-latency flash memory currently available, % we would like to note that
% PCIe Gen 6.0 with $N_{\max} = 15{,}360$ will greately relax this 3 usec limit.
% PCIe Gen 6.0 will increase the concurrency 20-fold to $N_{\max} = 15{,}360$ while quadrupling the bandwidth $W$.
% This means that most likely the latency will be limited by the GPU concurrency.
% Our current GPU concurrency is found to be 2,048 in our BFS experiment
% 4-fold increase in GPU concurrency from PCIe Gen 4.0 limit 768 will relax this 3 usec limit.
% ... this is too complicated and may give an impression that our assumption that the PCIe is the bottlneck will not apply in the near future (it will apply to the bandwidth but not to the # outstandings)
% In the meantime, CXL memory implementations will improve as the technology matures, which would potentially contribute to reducing the overall latency to leave room for longer-latency memory devices.
% \markchanges{CXL memory implementations will improve as the technology matures, contributing to reducing the overall latency while accommodating low-latency memory devices.}
\markchanges{we believe this requirement is within reach considering the technological advancements in memory devices as well as in CXL memory implementations.}

% and even more cost-effective memory devices such as flash having even longer latency than those modeled in our crude implementation may be used in the future.
% In PCIe Gen 6.0 with $N_{\max} = 15{,}360$, this situation is likely to change, which means even longer latency may be tolerated in the near future.

\vspace{0.5mm}
\noindent
{\bf Read-only workloads:}
\camerareadyrevision{The graph processing workloads evaluated in this paper are all read-only.
While CXL is a cache-coherent protocol, the coherency overhead should be minimal for read-only workloads,}
\additionalrevision{if any.}
% The overhead, if any, is unlikely to depend on the address alignement size $a$ either, as the CXL access size is 64 B irrespective of the alignment size used in the GPU.}
\camerareadyrevision{For workloads involving write access, there will be a number of additional factors to be considered, including cache coherency mentioned above and write characteristics of flash memory,}
\additionalrevision{all of which may have dependencies on the address alignment size $a$ and data transfer size $d$.
These factors can translate to performance impacts.}
\camerareadyrevision{Furthermore, additional care will have to be taken if memory persistency needs to be ensured.}

% However, write access and coherence are not considered.
% Given that CXL is cache-coherent, would different address alignment sizes impact the coherence overhead?
% The work finds that flash-based CXL memory can used for constructing cost-effective systems, but that may add overhead due to persistency if write access is considered.

\vspace{0.5mm}
\noindent
{\bf Other system configurations:}
\camerareadyrevision{The premise of our analysis and evaluation is that the GPU onboard memory is limited and benefits from external memory through the PCIe link.
This does not always hold:
one can opt to use GPUs having large memory (e.g., 80 GB \cite{NVidiaA100}) or bundle multiple of them to create an even larger pool of GPU memory.
Moreover, some emerging architectures integrate a CPU and GPU on the same chip, bringing the host DRAM closer to the GPU \cite{NVidiaGraceHopper,AMD-MI300}.
In both cases, HBM (High Bandwidth Memory) enables a much higher throughput than what a PCIe link offers.
If performance is more heavily weighted than cost, these system configurations can be compelling options.
In the meantime, our projection is that PCIe-attached GPUs with limited memory will continue to constitute cost-effective options, and we believe flash-based CXL memory can make those options even more appealing.}
% New CPU-GPU / APU designs like Grace Hopper and MI300 bring host memory much closer to the GPU. Should this potentially move the reference point against which future CXL-attached memory should be compared?
% The paper does not discuss HBM, and the GPUs used in the study use GDDR instead. With the latest GPUs allowing up to 80GB HBM2 and able to be connected to 4 others via NVLINK direct connections or more via NVLINK switches, this seems like a compelling option if performance is more heavily weighted than cost. It seems that this should at least be mentioned in the paper.

\camerareadyrevision{Currently, CXL memory access from the GPU goes through the CPU.}
\additionalrevision{The GPU does not use the CXL protocol, and issues requests in the same way as when accessing the host DRAM, and the CPU translates them into CXL access.}
\camerareadyrevision{However, future GPUs may implement the CXL interface to directly communicate with CXL memory. % over a PCIe link.
Still, our analysis will remain valid so far as the PCIe link to the GPU continues to be the bottleneck.
As the direct communication will reduce the CXL memory latency seen from the GPU, it will likely become easier to achieve a latency of a few microseconds.}

% Can GPU be directly connected to CXL memory through PCIe instead of through CPU and then CXL? The paper can add a paragraph justifying this design choice.

\vspace{0.5mm}
\noindent
{\bf Other graph formats and preprocessing:}
\additionalrevision{Our evaluation has used a BaM-like implementation for XLFDD, and EMOGI for CXL memory, both of which perform fine-grained access to external memory on demand.
We have used them because they achieve state-of-the-art runtime performance.
They are also beneficial in that the GPU can directly take an input graph in the standard CSR format without preprocessing.
However, fine-grained access implies a small average data transfer size $d$ over the PCIe link.
Although a larger $d$ will relax the latency and IOPS requirements for external memory, we cannot arbitrarily increase $d$ as it depends on the input graph: increasing $d$ beyond the average edge sublist size will increase the RAF, leading to a negative performance impact.
Therefore, in order to increase $d$ further, it would be interesting to consider tailored graph formats and preprocessing such as \cite{GraphReduce,Graphie,Subway}.}

\section{Related Work}

% We review related work in two categories: graph processing on external memory and CXL evaluation.

% \subsection{Graph Processing on External Memory}

% \subsubsection{CPU graph processing}
{\bf CPU graph processing:}
When graph data fits in the host DRAM, in-memory graph processing methods like Galois \cite{Galois} and GAP \cite{GAP2015} can be used.
When it does not fit in the DRAM, one option is distributed processing \cite{S38}, but communication overheads limit the performance \cite{S22,S39}.
For this reason, single-node approaches utilizing storage devices as external memory have been proposed \cite{S35,S48,S37,S27,S28,S43,S31}.
Graphene \cite{S35} achieves excellent performance close to in-memory solutions for algorithms mainly involving sequential access such as PageRank, but it is significantly slower if random access is required like in BFS.
Graph algorithms involving random access has been shown to approach in-memory speeds if executed on low-latency flash-based storage by accessing it using lightweight context switching to hide latency \cite{VLDB-suzuki}.
In this paper we address random access graph workloads on the GPU using external memory, and find quite different requirements than the CPU case.

% \subsubsection{GPU graph processing on the host DRAM}
\vspace{0.5mm}
\noindent
{\bf GPU graph processing on the host DRAM:}
It is natural to consider taking advantage of massive compute resources of the GPU for graph processing.
As the GPU onboard memory is even more
limited, many prior works propose to place graph data on the host DRAM \cite{E25,E19,E39,E37,E23}
(in contrast to the CPU case, the host DRAM is viewed as external memory from the GPU).
These methods are based on a unified virtual memory (UVM) approach where portions of the host DRAM are copied to the GPU memory via paging at a 4 kB granularity \cite{NVidiaUVM}.
EMOGI instead uses zero-copy access and has shown that this fine-grained direct access significantly reduces the RAF compared with the UVM approach \cite{EMOGI2020}.
This paper has shown that EMOGI stays as performant even if the external memory latency is longer than the host DRAM, up to a few microseconds. 
% {\bf Host-memory based graph processing on GPU.}
% \cite{EMOGI2020} employs zero-copy based host-memory access rather than using UVM\cite{NVidiaUVM}, which is used in many prior works \cite{E25,E19,E39,E37,E23} and others referred in \cite{EMOGI2020}.
% Compared to these UVM based approaches, \cite{EMOGI2020} successfully reduce the read-amplification caused by UVM.
% In that sense, our method of small-sized direct-access follows the idea of \cite{EMOGI2020} and expands it to storage access.

% It has been shown that this fine-grained direct access significantly reduces the RAF compared with a more prevalent, unified virtual memory (UVM) approach where portions of the host DRAM are copied to the GPU memory via paging at a 4 kB granularity.

%\subsubsection{GPU graph processing on storage}
\vspace{0.5mm}
\noindent
{\bf GPU graph processing on storage:}
BaM introduced a first GPU-centric storage access method that does not involve CPU intervention \cite{BaM2023}.
While there are several prior works in GPU-centric approaches \cite{BaM41,BaM43,BaM30,BaM49,BaM34}, they rely on the CPU to handle storage access and use the GPU memory as a staging buffer for their data transfer.
BaM has shown that it achieves competitive runtimes with EMOGI when the EMOGI's runtimes include the time for loading graph data from SSDs.
This paper has shown that, by using external memory based on microsecond-latency flash memory, we can achieve even faster runtimes so that they are close to those of EMOGI even if we exclude EMOGI's file loading time.

% \noindent{\bf Hardware extensions.}
% Hardware support for storage-class memories from GPU are proposed in \cite{BaM50,BaM51,BaM54,BaM55,BaM56,BaM13,BaM16},
% as well as storage with memory-interface is announced from flash-memory vendors \cite{MSSSD,CXLXL}.
% The proposed ideas of direct access and/or fine-grain accesses from GPU will work with these approaches
% as discussed in Sec.~\ref{sec:SSDs-with-CXL-memory-interface}.

\vspace{0.5mm}
\noindent
{\bf CXL analysis and evaluation:}
CXL is an emerging standard that is attracting attention not only from industry but also from research communities.
Analysis and evaluation of CXL-enabled systems are being conducted ranging from memory pooling in general \cite{Pond,CXLMemPool,DemystifyCXL,CXLMemSim,DirectCXL}, to more specific applications such as
% failure-tolerant
machine learning \cite{TrainingCXL} and in-memory databases % management systems
\cite{Elastic}.
CXL studies involving accelerators such as GPU and FPGA are appearing \cite{TrainingCXL,CXL-GPU-FPGA}.
Our work complements these studies and deals with GPU graph processing on CXL memory for the first time.

% TrainingCXL implements CXL on FPGA

\section{Conclusion}

We have presented analysis and evaluation of GPU graph traversal using CXL-based external memory.
Given the nature of the workload where on-demand, fine-grained random reads are bottlenecked by the PCIe link to the GPU, we note that a small address alignment of around 32 B, along with appropriately-sized data transfer close to the average edge sublist size of a few hundred bytes, will lead to an optimal runtime.
This translates into the requirements for external memory, which are random read performance of a few hundred MIOPS and a latency of a few microseconds, suggesting the possibility that CXL memory with longer latency, including that equipped with low-latency flash memory, may be used as external memory to achieve performance comparable to the host DRAM.

To support these observations, we have conducted evaluation using two FPGA-based external memory prototypes, one is a storage device with low-latency flash memory and the other DRAM-based CXL memory with adjustable latency, and we have demonstrated GPU graph processing speeds close to using the host DRAM when the external memory latency is under a few microseconds.

While our evaluation is limited by the current availability of CXL devices, we believe it provides first preliminary characterizations of GPU access to CXL memory, which we hope leads to insights into how cost-effective systems may be constructed potentially by incorporating flash-based CXL memory.

%\begin{acks}
%  We thank the anonymous reviewers for thier valuable feedback, which helped us strengthen the paper.
%\end{acks}

\appendix
\section{Latency Bridge Design}
\label{sec:latency_bridge}
\begin{figure}[h]
  \centering
  \includegraphics[trim=0 260 165 0,clip,width=\columnwidth]{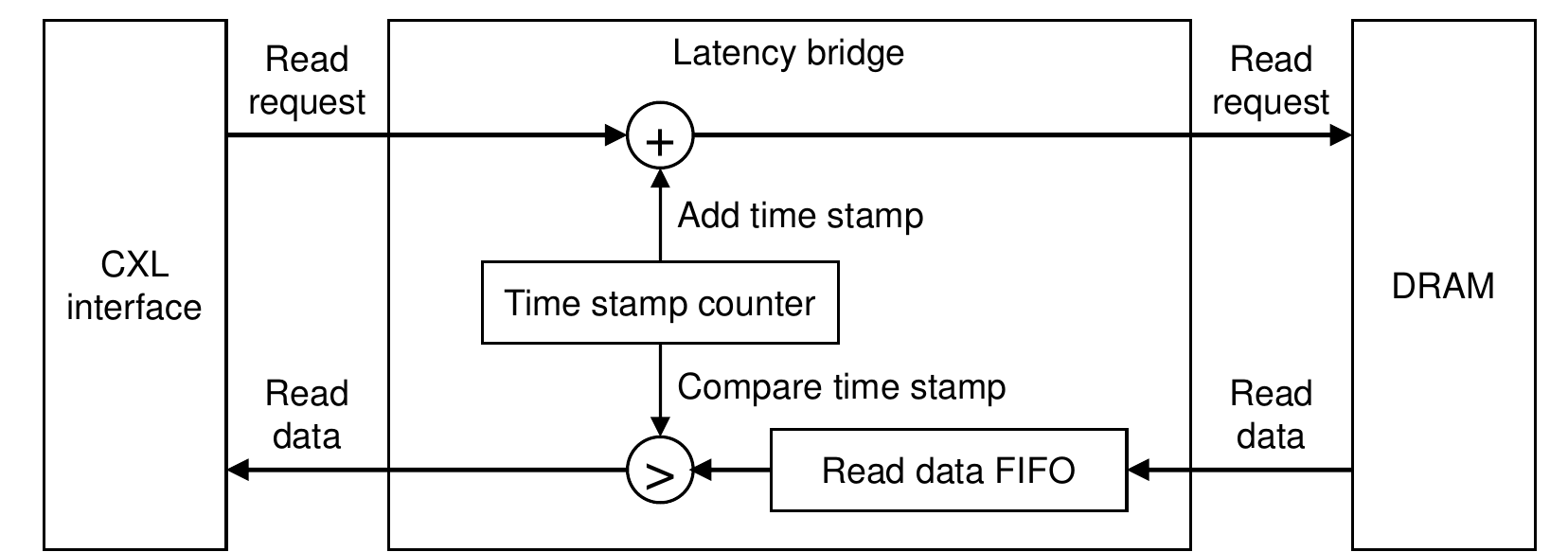}
  \caption{Block diagram of the latency bridge.}
  \label{fig:latency_bridge}
\end{figure}

\camerareadyrevision{The latency bridge described in Section~\ref{sec:cxl_imple} is implemented as shown in Figure~\ref{fig:latency_bridge}.
We add a time stamp to an incoming read request, read data from the DRAM, and push it to a FIFO along with the time stamp.
% The read data is pushed to a FIFO with its associated time stamp.
When the current time becomes greater than the time stamp of the FIFO head by a specified additional latency, the data is popped and sent to the CPU through the CXL interface.
As the CXL interface of Intel Agilex\textregistered 7 FPGA processes requests in order at the time of this work, a FIFO is sufficient, but a slightly more involved design would be required to support out-of-order access.}

\section{Pointer Chasing from the GPU}
\label{sec:pointer_chasing}
\camerareadyrevision{We perform pointer chasing to measure the latency of CXL memory (and the host DRAM) from the GPU as described in Section~\ref{sec:cxl_character}.
In preparation, we allocate a \additionalrevision{16-GB} block of CXL memory and fill it with \additionalrevision{134} million 128-B indices (or pointers) each pointing to the next address to look at.
We run a single GPU warp \cite{CUDAwarps} to chase them: it reads the first pointer, then reads the next pointer stored at the address pointed to by the first pointer, and so on.
The pointers are set in such a way that the GPU has to move randomly in the \additionalrevision{16-GB} space.
The 32 GPU threads in a warp each fetch 4 B of a 128-B pointer and synchronize before reading the next pointer.
The runtime of this operation is determined by the memory latency as the next pointer is only available after reading the current pointer.}

%\Urlmuskip=0mu plus 1mu
\bibliographystyle{ACM-Reference-Format}
\bibliography{xlgpu}

% \appendix

\end{document}